\documentstyle[psfig]{mn}

\newif\ifAMStwofonts

\ifoldfss
  \newcommand{\rmn}[1] {{\rm #1}}

  \ifCUPmtlplainloaded \else
    \NewTextAlphabet{textbfit} {cmbxti10} {}
    \NewTextAlphabet{textbfss} {cmssbx10} {}
    \NewMathAlphabet{mathbfit} {cmbxti10} {} 
    \NewMathAlphabet{mathbfss} {cmssbx10} {} 
  \fi
  \ifAMStwofonts
    \ifCUPmtlplainloaded \else
      \NewSymbolFont{upmath} {eurm10}
      \NewSymbolFont{AMSa} {msam10}
      \NewMathSymbol{\upi}     {0}{upmath}{19}
      \NewMathSymbol{\umu}     {0}{upmath}{16}
      \NewMathSymbol{\upartial}{0}{upmath}{40}
      \NewMathSymbol{\leqslant}{3}{AMSa}{36}
      \NewMathSymbol{\geqslant}{3}{AMSa}{3E}

    \fi
  \fi
\fi 

\ifnfssone
  \newmathalphabet{\mathit}
  \addtoversion{normal}{\mathit}{cmr}{m}{it}
  \addtoversion{bold}{\mathit}{cmr}{bx}{it}
  \newcommand{\rmn}[1] {\mathrm{#1}}

  \def\textbfit{\protect\txtbfit}
  
  \long\def\txtbfit#1{{\fontfamily{cmr}\fontseries{bx}\fontshape{it}%
    \selectfont #1}}

  \newmathalphabet{\mathbfit} 
  \addtoversion{normal}{\mathbfit}{cmr}{bx}{it}
  \addtoversion{bold}{\mathbfit}{cmr}{bx}{it}
  \newmathalphabet{\mathbfss} 
  \addtoversion{normal}{\mathbfss}{cmss}{bx}{n}
  \addtoversion{bold}{\mathbfss}{cmss}{bx}{n}
  \ifAMStwofonts
    \ifCUPmtlplainloaded \else
      \UseAMStwoboldmath
      \makeatletter
      \new@mathgroup\upmath@group
      \define@mathgroup\mv@normal\upmath@group{eur}{m}{n}
      \define@mathgroup\mv@bold\upmath@group{eur}{b}{n}
      \edef\UPM{\hexnumber\upmath@group}
      \new@mathgroup\amsa@group
      \define@mathgroup\mv@normal\amsa@group{msa}{m}{n}
      \define@mathgroup\mv@bold\amsa@group{msa}{m}{n}
      \edef\AMSa{\hexnumber\amsa@group}
      \makeatother
      \mathchardef\upi="0\UPM19
      \mathchardef\umu="0\UPM16
      \mathchardef\upartial="0\UPM40
      \mathchardef\leqslant="3\AMSa36
      \mathchardef\geqslant="3\AMSa3E
    \fi
  \fi
\fi 

\ifnfsstwo
  \newcommand{\rmn}[1] {\mathrm{#1}}

  \def\textbfit{\protect\txtbfit}
  
  \long\def\txtbfit#1{{\fontfamily{cmr}\fontseries{bx}\fontshape{it}%
    \selectfont #1}}

  \DeclareMathAlphabet{\mathbfit}{OT1}{cmr}{bx}{it}
  \SetMathAlphabet\mathbfit{bold}{OT1}{cmr}{bx}{it}
  \DeclareMathAlphabet{\mathbfss}{OT1}{cmss}{bx}{n}
  \SetMathAlphabet\mathbfss{bold}{OT1}{cmss}{bx}{n}
  \ifAMStwofonts
    \ifCUPmtlplainloaded \else
      \DeclareSymbolFont{UPM}{U}{eur}{m}{n}
      \SetSymbolFont{UPM}{bold}{U}{eur}{b}{n}
      \DeclareSymbolFont{AMSa}{U}{msa}{m}{n}
      \DeclareMathSymbol{\upi}{0}{UPM}{"19}
      \DeclareMathSymbol{\umu}{0}{UPM}{"16}
      \DeclareMathSymbol{\upartial}{0}{UPM}{"40}
      \DeclareMathSymbol{\leqslant}{3}{AMSa}{"36}
      \DeclareMathSymbol{\geqslant}{3}{AMSa}{"3E}
    \fi
  \fi
\fi 
\ifCUPmtlplainloaded \else
  \ifAMStwofonts \else 
    \def\upi{\pi}
    \def\umu{\mu}
    \def\upartial{\partial}
  \fi
\fi

\title[ROSAT HRI observations of M13 and M92]
   {\textbfit{ROSAT\/} HRI observations of the globular clusters M13
   and M92} 
\author[D. Fox et al.]
   {D.~Fox,$^1$ W.~Lewin,$^1$ B.~Margon,$^2$ 
     J.~van~Paradijs,$^{3,4}$ F.~Verbunt$^5$\\
     $^1$MIT Center for Space Research, 77 Mass.\ Ave \#37-627,
   Cambridge, MA  02139-4307, USA\\ 
     $^2$Astronomy Department, University of Washington, Box 351580,
   Seattle, WA 98195-1580, USA \\
     $^3$Astronomical Institute `Anton Pannekoek' and Center for
   High-Energy Physics, Kruislaan 403, 1098 SJ Amsterdam, The
   Netherlands \\
     $^4$University of Alabama in Huntsville, Huntsville, AL 35899,
   USA \\ 
     $^5$Astronomical Institute, Utrecht University, P.O. Box 80000,
   NL-3508 TA Utrecht, The Netherlands}
\date{Accepted date. Received date.}
\pagerange{\pageref{firstpage}--\pageref{lastpage}}
\pubyear{1996}

\newcommand{\einstein}{{\it Einstein\/}}
\newcommand{\rosat}{{\it ROSAT\/}}
\newcommand{\ergcms}{erg cm$^{-2}$ sec$^{-1}$}
\newcommand{\ergsec}{erg sec$^{-1}$}
\newcommand{\thsp}{$\,$}
\newcommand{\mavg}{\mbox{$\langle m \rangle$}}
\newcommand{\Lx}{\mbox{$L_{\rm x}$}}
\newcommand{\Rc}{\mbox{$R_{\rm c}$}}
\newcommand{\Msol}{\mbox{$M_\odot$}}
\newcommand{\Rext}{\mbox{$R_{\rm ext}$}}
\newcommand{\Pext}{\mbox{$P_{\rm ext}$}}
\newcommand{\alphaRA}{\mbox{$\alpha_{\rm 2000}$}}
\newcommand{\deltaDEC}{\mbox{$\delta_{\rm 2000}$}}

\newcommand{\cmsqr}{\mbox{cm$^{-2}$}}
\newcommand{\rah}{$^{\rm h}$}
\newcommand{\ram}{$^{\rm m}$}

\begin{document}
\label{firstpage}
\maketitle
\begin{abstract}
  We report on 40 kiloseconds of \rosat\ HRI observations of the
  globular cluster M92 (NGC 6341) and 20 kiloseconds of observations
  of the globular cluster M13 (NGC 6205).  We find that the
  low-luminosity ($10^{32.5}$ \ergsec\ at 7.5 kpc) source previously
  observed near the core of M92 with the \rosat\ PSPC remains
  unresolved at HRI resolution; we can identify it with M92 with 99
  per cent confidence.  In M13 we find that the source seen with the
  \rosat\ PSPC lies within the core and is possibly associated with
  the cluster (96 per cent confidence).  We find probabilities of 99.8
  per cent (M92) and 98 per cent (M13) for the presence of additional
  unresolved emission within these globular clusters.  We interpret
  these results in light of current theories regarding the
  low-luminosity sources.
\end{abstract}



\section{Introduction}
\label{sec:intro}
The low-luminosity X-ray sources ($\Lx < 10^{34.5}$ \ergsec) in
globular clusters were first distinguished in the \einstein\ 
globular-cluster survey of Hertz \& Grindlay (1983a,b; hereafter HG83a
and HG83b).  Separated by 1.5 orders of magnitude in flux from the
high-luminosity sources that are convincingly explained as low-mass
X-ray binaries (LMXBs; see Lewin, Van Paradijs \& Taam 1993 for a
review), they seemed a new population, and over a decade of
observations has confirmed this view.   

Recent optical and ultraviolet (UV) globular cluster observations have
yielded interesting objects in the cluster cores, and these have been
suggested as counterparts for the low-luminosity X-ray sources.  The
ultraviolet variables V1 and V2 detected in the core of 47~Tuc
\cite{Paresceet92,ParDeM94} are a case in point.  But since the cores
of clusters are crowded with interesting objects, positional
coincidence is not a convincing argument for association unless it is
accompanied by corresponding temporal variability or by a spatial
pattern present in both X-rays and the optical/UV.  The latter is the
case in the core of NGC 6397, where three low-luminosity sources have
been identified with three H$\alpha$ emission objects that are
probably cataclysmic variables (CVs; Cool et al.\ 1995a, Grindlay et
al.\ 1995).  

These counterparts notwithstanding, the nature of the full source
population remains uncertain.  Some of the initial HG83b sources are
almost certainly foreground or background objects rather than
low-luminosity sources \cite{MB87,Verbuntet95,Coolet95b}, while
proposals for the population itself include CVs (HG83a), LMXBs --
either in quiescence \cite{Verbuntet84,VanPet87} or viewed at high
inclination \cite{WhiteMason85} -- millisecond pulsars
\cite{Danneret94}, and RS CVn systems \cite{Bailynet90,Verbuntet93}.

With the addition of the data reported here, a total of 38 galactic
globular clusters have been subjected to deep imaging with \rosat\ 
(HRI and PSPC).  Low-luminosity sources have been seen within or near
the cores of 22 of these clusters; perhaps as many as 37 such sources
have now been detected (see Johnston \& Verbunt 1996 for a full
discussion).  The present observations aim to constrain this
population further, taking advantage of the HRI's high resolution
(FWHM $\sim$ 7 arcsec) to evaluate the possible multiplicity of the
sources seen in PSPC observations of M92 and M13
\cite{HJV94,Margonet94a} and to search for extended emission from the
cores of these clusters.


\section{Observations and Data Reduction}
\label{sec:observe}
Observations were made with the \rosat\ X-ray telescope
\cite{Trumperet91} and High Resolution Imager (HRI; David et al.\ 
1995), in pointing mode.  The initial observation of M92 (17.9 ksec
live time) took place during 4--8 April 1994; the observation of M13
which was to follow was aborted for instrumental reasons and
rescheduled for the fall.  At the same time, 20 kiloseconds were
inserted into the schedule for a repeat observation of M92, so that
from 29 August to 5 September 1994 our entire original proposed
observation sequence was carried out (21.4 ksec live time on M13 and
another 20.0 ksec on M92).  Thus our total exposure of M92 is 37.9
ksec.

The data reduction began with processing by the \rosat\ Standard
Analysis Software System (SASS).  This software uses a series of
sliding cell passes over the image -- with detect cells ranging from
24 to 240 pixels on a side -- to find rough source positions, and
then refines the positions using a maximum-likelihood method.  At each
cell size, the source detection threshold is set so that one spurious
source detection is expected over the field of view.  Sources that
were detected at more than one cell size here were passed on to the
next stage of analysis.

At this point photon lists were created for the near vicinity ($\sim$
one arcmin) of each of the candidate sources.  With custom software we
determined source positions by a maximum-likelihood fit to the photon
distribution, using the canonical form of the HRI Point-Response
Function (PRF) \cite{Davidet95}.  Positional uncertainties due to the
finite-photon statistics of each source were determined by Monte Carlo
analysis and found to be about one arcsec for on-axis sources,
significantly less than the systematic uncertainty in the satellite
pointing ($\sim 5$ arcsec).

Using these positions, we determined source fluxes by a local
background subtraction: counts from an annulus (20--30 arcsec for
on-axis sources) about the source were scaled by area and subtracted
from the counts in the source disk (radius 17.5 arcsec on-axis).  Disk
and annuli radii were inflated off-axis to compensate for the
broadening of the HRI PRF -- at a minimum, allowing the source disk to
encompass 90 per cent of the source counts.  Count rates were
corrected for vignetting and converted to 0.1--2.4~keV fluxes,
assuming a 1~keV bremsstrahlung spectrum \cite{Margonet94b}\ and a
hydrogen column density of $10^{20.1}$ cm$^{-2}$.  The column
densities were determined by the cluster reddenings ($E_{B-V}$ = 0.02
for both M92 and M13, Peterson 1993, gives $A_V$ = 0.06, Zombeck 1990)
through the prescription of Predehl \& Schmitt (1995):
\[
    N_{\rm H} = 1.79 \times 10^{21} \, A_V \, {\rm cm}^{-2}. 
\]
If the actual source spectra are harder than assumed, our derived
source fluxes are hardly affected; if they are significantly softer,
this has a small effect (fluxes increase by 10 per cent for a 0.50~keV
temperature).

Source significances were determined by summing counts within a
critical radius -- selected by hand for each source -- and evaluating
the probability of the Poisson fluctuation above background.  The
quoted confidence level, then, refers to the probability of not seeing
such a fluctuation (over an area of that size) {\em anywhere\/} in the
field of view.  Sources determined to be of less than $3\sigma$
significance (99.7 per cent confidence) at this point were eliminated.

We analysed the two M92 data sets in tandem: the photons in areas
surrounding preliminary source detections in either dataset were
output from both and subjected to separate maximum-likelihood
analyses.  Using the positions of the four strongest sources in the
field of view -- sources A--D, which had firm detections in both
data sets -- we rotated (by 0.14 degrees) and shifted (by 2.3 arcsec)
the second observation to closest alignment with the first.  We then
performed the same analysis on the photons extracted from the areas in
the summed image, which provided our canonical positions, fluxes and
errors.  The separate datasets were then compared for evidence of
source variability.

The parameters of the observed sources are presented in
Tables~\ref{tbl:m92obs} and \ref{tbl:m13obs}; Table~\ref{tbl:m92var}
\begin{table*}
\begin{minipage}{155mm}
\caption{Sources in the field of view of our M92 observations (37925
  seconds live time).}
\begin{tabular}{ll@{}l@{}ll@{}l@{}lrr@{$\pm$}rrrrrr}
  M92 & \multicolumn{3}{c}{\alphaRA} & \multicolumn{3}{c}{\deltaDEC} &
  $\Delta$ (\arcmin)& \multicolumn{2}{c}{Counts} & HRI Flux  
    & $\sigma$ & PSPC Flux & Simbad & $\Delta_{\rm Sim}$ (\arcsec)\\ 
  \hline
  A & 17\rah\thsp&16\ram\thsp&31\fs 9 & 43\degr\thsp&02\arcmin\thsp&37\arcsec 
    & 8.5  & $1128$ & $10$ & $116.0 \pm 3.9$ 
    & 121. & $131.5 \pm 2.0$ & JVH 10 & 5.9 \\
  B & 17&17&00.4 & 43&10&51 & 3.0 & $  70$ & $ 6$ & $  7.2 \pm 1.3$ 
    &  8.7 & $  6.7 \pm 0.5$ & -- &  \\
  C & 17&17&06.3 & 43&08&23 & 0.3 & $  47$ & $ 7$ & $  4.9 \pm 1.4$ 
    &  6.5 & $ 11.8 \pm 0.7$ & -- &  \\
  D & 17&18&09.7 & 43&09&06 & 11.4 & $ 66$ & $ 7$ & $  6.8 \pm 1.5$ 
    &  5.9 & $  3.1 \pm 0.4$ & JVH 07 & 3.3 \\
  E & 17&16&37.3 & 43&12&25 & 6.9 & $  45$ & $ 7$ & $  4.6 \pm 1.3$ 
    &  4.4 & $<$1.0 & V798 Her & 0.2 \\
  F & 17&17&29.5 & 43&19&49 & 12.3 & $ 51$ & $ 6$ & $  5.3 \pm 1.4$ 
    &  4.5 & $ 10.6 \pm 0.7$ & JVH 02 & 7.2 \\
  G & 17&16&48.4 & 42&55&01 & 13.6 & $ 79$ & $12$ & $  8.1 \pm 2.2$ 
    &  7.2 & $  3.3 \pm 0.4$ & JVH 12 & 6.7 \\
 \hline
\end{tabular}
\medskip

$\Delta$ is the distance of the source from the cluster centre at
17\rah 17\ram 07\fs 3 +43\degr 08\arcmin 11\arcsec (Trager, Djorgovski
\& King 1993).  Fluxes (0.1--2.4 keV) are in units of $10^{-14}$
\ergcms, assuming a 1~keV bremhsstrahlung spectrum and a column
density of $10^{20.1}$ \cmsqr.  Quoted significances $\sigma$ refer to
the probability of not finding a background fluctuation of that
magnitude anywhere in the field of view (the ratio of the observed
flux to its error may well be smaller). $\Delta_{\rm Sim}$ is the
distance of the Simbad object from the corresponding HRI source.
\label{tbl:m92obs}
\end{minipage}
\end{table*}
\begin{table*}
\begin{minipage}{120mm}
\caption{Sources in the field of view of our M13 observation (21378
  seconds live time).}  
\begin{tabular}{ll@{}l@{}ll@{}l@{}lrr@{$\pm$}rrrr}
  M13 & \multicolumn{3}{c}{\alphaRA} & \multicolumn{3}{c}{\deltaDEC} &
  $\Delta$ (\arcmin)& \multicolumn{2}{c}{Counts} & HRI Flux & $\sigma$ 
  & PSPC Flux  \\ 
 \hline
 A & 16\rah\thsp& 40\ram\thsp& 32\fs 1 & 36\degr\thsp&23\arcmin\thsp&23\arcsec & 14.6 & $ 68$ & $11$ & $12.3 \pm 3.4$ & 3.8 & $ 2.8 \pm 0.4$  \\
  B & 16&40&56.3 & 36&34&07 & 11.2 & $178$ & $ 6$ & $32.4 \pm 3.1$ 
    &18.6 & $<$1.0 \\
  C & 16&40&58.5 & 36&24&30 &  9.2 & $ 35$ & $ 5$ & $ 6.4 \pm 1.8$ 
    & 6.6 & $ 2.7 \pm 0.4$ \\
  D & 16&41&11.1 & 36&32&33 &  7.8 & $114$ & $ 5$ & $20.8 \pm 2.4$ 
    &12.5 & $32.2 \pm 1.0$ \\
  E & 16&41&18.6 & 36&26&48 &  4.7 & $ 26$ & $ 5$ & $ 4.8 \pm 1.7$ 
    & 4.0 & $ 2.5 \pm 0.4$ \\
  G & 16&41&44.0 & 36&27&59 &  0.6 & $ 58$ & $ 5$ & $10.5 \pm 2.0$ 
    & 7.9 & $ 9.7 \pm 0.6$ \\
  H & 16&41&56.6 & 36&35&16 &  8.2 & $ 27$ & $ 5$ & $ 4.8 \pm 1.7$ 
    & 3.2 & $ 5.5 \pm 0.5$ \\
  I & 16&42&03.6 & 36&24&15 &  5.6 & $ 29$ & $ 4$ & $ 5.3 \pm 1.6$ 
    & 5.1 & $ 3.5 \pm 0.4$ \\
  J & 16&42&19.8 & 36&31&53 &  8.8 & $ 38$ & $ 5$ & $ 7.0 \pm 1.9$ 
    & 5.8 & $<$1.0 \\
  N & 16&41&52.8 & 36&40&23 & 13.0 & $ 42$ & $ 7$ & $ 7.7 \pm 2.3$ 
    & 3.3 & $<$1.0 \\
  P & 16&41&59.7 & 36&44&21 & 17.1 & $ 61$ & $14$ & $11.1 \pm 4.1$ 
    & 3.3 & $<$1.0 \\
  Q & 16&42&35.7 & 36&18&13 & 14.4 & $ 56$ & $10$ & $10.3 \pm 3.1$ 
    & 4.1 & $ 3.3 \pm 0.4$ \\
  \hline
\end{tabular}
\medskip

$\Delta$ is the distance of the source from the cluster centre at
16\rah 41\ram 41\fs 5 +36\degr 27\arcmin 37\arcsec (Trager et al.\ 
1993).  Fluxes (0.1--2.4 keV) are in units of $10^{-14}$ \ergcms,
assuming a 1~keV bremhsstrahlung spectrum and a column density of
$10^{20.1}$ \cmsqr.  Quoted significances refer to the probability of
not finding a background fluctuation of that magnitude anywhere in the
field of view (the ratio of the observed flux to its error may well be
smaller).  No Simbad counterparts were found within one arcminute of
any source.
\label{tbl:m13obs}
\end{minipage}
\end{table*}
\begin{table}
\caption{Comparison of the two observations of M92.}
\begin{tabular}{lrrrrr}
  M92 & Flux 1      & Flux 2         & Sig 1 & Sig 2 & Variable? \\ \hline
  A  & $104.5 \pm 5.3$ & $ 127. \pm 5.6$ & 72.3 & 73.0 & 3.2$\sigma$ \\
  B  & $  7.9 \pm 1.9$ & $  7.8 \pm 1.9$ &  6.4 &  6.0 & -- \\
  C  & $  4.3 \pm 1.9$ & $  6.8 \pm 1.9$ &  3.3 &  5.1 & -- \\
  D  & $  6.2 \pm 2.0$ & $  5.6 \pm 2.1$ &  4.5 &  3.8 & -- \\
  E  & $  5.5 \pm 1.8$ & $  6.1 \pm 2.0$ &  4.4 &  4.5 & -- \\
  F  & $  3.5 \pm 1.9$ & $  6.9 \pm 2.0$ &  2.7 &  4.9 & 2.3$\sigma$ \\
  G  & $  2.2 \pm 2.8$ & $ 12.5 \pm 3.4$ &  1.4 &  5.1 & 5.8$\sigma$ \\
  \hline
\end{tabular}
\medskip

Fluxes (0.1--2.4 keV) are in units of $10^{-14}$ \ergcms.  ``Flux~1''
and ``Flux~2'' are the fluxes observed in our two observations, and
``Sig~1'' and ``Sig~2'' the corresponding significances.  The
significances here refer to the probability of not seeing such a
fluctuation above background within the source disk only (17.5 arcsec
radius on-axis).  Sources variable at greater than the two-sigma level
are indicated.
\label{tbl:m92var}
\end{table}
presents a comparison of the sources present in the two observations
of M92.  The quasar QSO~1715+432 \cite{Harriset92}, present in the M92
field of view, was not detected; no quasars within the M13 field of
view were found in the catalog of Hewitt \& Burbidge 1993.  The four
1.4~GHz radio sources within the M13 core (including one pulsar)
observed by Johnston, Kulkarni \& Goss (1991) were not detected.

Our limiting fluxes for detection of sources on-axis were $2.9 \times
10^{-14}$ and $4.7 \times 10^{-14}$ \ergcms, respectively, for M92 and
M13 (corresponding, in turn, to X-ray luminosities of $2.0 \times
10^{32}$ and $2.9 \times 10^{32}$ \ergsec\ at the cluster distances of
7.5 and 7.2 kpc), where again, a 1~keV bremsstrahlung spectrum has
been assumed.  To evaluate whether we see an overdensity of sources in
these fields of view, we first compare the number of sources seen to
that expected from the $\log N$--$\log S$ relation of \rosat\ deep
surveys (Hasinger et al.\ 1993; hereafter Ha93).  For purposes of this
comparison, we assume an $\alpha=1$ power-law spectrum for the
background sources; this is roughly the best-fit spectrum given by
Ha93.  Adopting this spectrum increases the calculated source fluxes
by 10 per cent: our limiting fluxes become $3.2 \times 10^{-14}$ and
$5.3 \times 10^{-14}$ \ergcms\ for on-axis sources in M92 and M13.
Because of vignetting and the broadening of the PRF, these limiting
fluxes increase off-axis, but we ignore that effect here.

Using the soft X-ray $\log N$--$\log S$ from Ha93 and converting from
their smaller bandpass (0.5--2 keV), we find that at these limiting
fluxes we expect $37.0 \pm 3.5$ and $16.5 \pm 1.4$ background sources
per square degree for our M92 and M13 observations, respectively.
Adding in the Poisson uncertainties on these numbers, we expect to
find $13.7 \pm 3.9$ and $6.1 \pm 2.5$ sources in the M92 and M13
fields of view.  We see seven sources in the M92 field of view and
twelve sources in the M13 field of view, which is consistent with
these predictions.

As a further check, we compare our observations to the PSPC
observations of the same clusters \cite{Margonet94b}.  Within the HRI
field of view for M92 we find ten PSPC sources above the HRI flux
limit, consistent -- if we allow for modest source variability -- with
the seven observed.  Similarly, in the HRI field of view for M13 we
find 13 PSPC sources above the HRI flux limit, consistent with the
twelve observed.  Tables~\ref{tbl:m92obs} and \ref{tbl:m13obs} show
that six of our seven M92 HRI sources, and eight of our twelve M13 HRI
sources, were also seen in the PSPC observations.

A query of the Simbad database returned a list of identified galactic
objects within our fields of view, and when these fell close to our
sources, they are also included in the Tables.  The coincidences in
the M92 field of view are striking: all five of our sources outside
the M92 core region have a Simbad counterpart foreground star within
8 arcsec.  We have taken advantage of these coincidences to adjust our
aspect for that observation.  

No Simbad counterparts were found for our M13 sources; however,
examination of a Palomar Optical Sky Survey image of M13 from the
STScI Digitized Sky Survey reveals probable bright but anonymous
optical counterparts to the sources M13B and M13D.  It also reveals
that the optical field is too crowded to allow firm counterpart
identifications for these sources or for sources M13E, M13G and M13I.
In all, then, our M13 observation finds seven sources outside the M13
core region without optical counterparts, where our M92 observation
finds counterparts for every source.  We postulate no underlying
reason for this discrepancy.


\section{Discussion: Core Sources}
\label{sec:cores}

\begin{figure}
 \begin{center}
   ~\psfig{figure=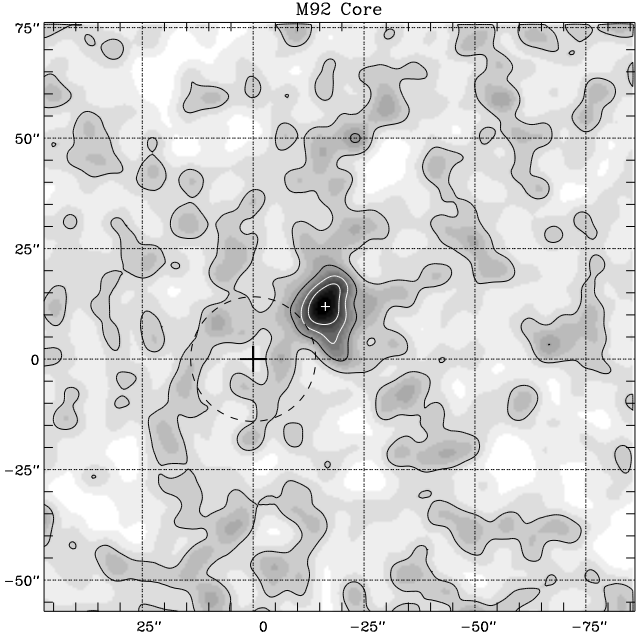,bbllx=77pt,bblly=228pt,bburx=554pt,bbury=699pt,angle=90,width=8.3cm}~
 \end{center}
 \caption{The core of M92.  This image (smoothed with the HRI PRF;
   north is up and east is to the left) shows our summed exposure of
   the core region of M92. The large black cross indicates the centre
   of the cluster and the systematic uncertainty in our (corrected)
   pointing.  The dashed line indicates the extent of the cluster
   core, and the small white cross, our best position for the central
   X-ray source and its statistical uncertainty.  Contours are at
   multiples of the background count rate.  The elliptical blurring
   along a northwest-southeast axis, present in our second
   observation, can be seen to distort the source profile of M92C
   here.}
 \label{fig:m92core}
\end{figure}

\begin{figure}
 \begin{center}
   ~\psfig{figure=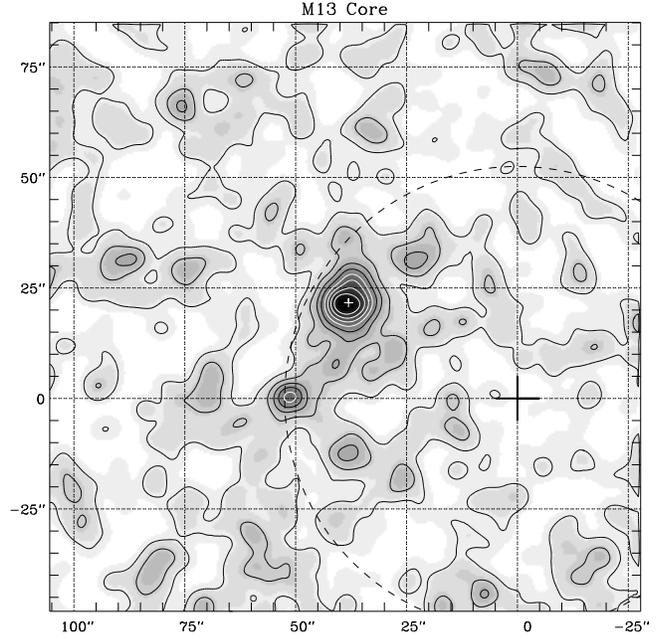,bbllx=77pt,bblly=228pt,bburx=554pt,bbury=699pt,angle=90,width=8.3cm}~
 \end{center}
 \caption{The core of M13 (see Fig.~\protect\ref{fig:m92core} for
   an explanation of the figure elements); north is up and east is to
   the left.  The bump to the southeast of M13G -- a 3.5$\sigma$ local
   fluctuation -- is not significant (we expect 50 such fluctuations
   over the field of view).  The distortion from a single-source
   profile mentioned in the text is evident in the bump immediately
   north of the cross marking Source~G.  }
 \label{fig:m13core}
\end{figure}

One source in each field of view (M92C and M13G; see
Figs.~\ref{fig:m92core} and \ref{fig:m13core}) is close enough in
projection to the centre of its cluster to warrant further analysis:
the a posteriori chances of finding a field source anywhere closer
than {\it twice\/} the distance of sources M92C and M13G from their
cluster centres are only 1.1 per cent and 5 per cent, respectively.

To construct a credible statistic for the significance of the
association, we make a comparison of two probability densities,
evaluated at the source's observed position: on the one hand, the
probability density for unassociated field sources (roughly constant
across the field of view), and on the other hand, the probability
density at that point for a binary system within the globular cluster.
For the second value we use formula (11) of Lightman, Hertz \&
Grindlay (1980) for the projected probability density of a test mass
greater than the average mass $\mavg\approx 0.6\Msol$ of a cluster's
core population:
\[
     p(\bmath{x}) \, d^2 x  =  
           \frac{3}{2\pi} (q-1)(1+x^2)^{-(3q-1)/2} \, d^2 x,
\]
where $\bmath{x}$ is the vector position of the source -- relative to
the cluster centre -- in units of the cluster core radius, and $q
\equiv m_{\rm x}/\mavg$ is the ratio of the test mass to \mavg.  (This
formula should properly be convolved with our uncertainties, but we
neglect that step.)  We take a $q$ value of 2.3 ($m_{\rm x} =
1.4\Msol$) -- average for a CV system in the clusters -- as being
reasonable for the low-luminosity sources.  The results of this
calculation are shown in Table~\ref{tbl:core}: we find a probability
\begin{table}
\caption{Sources close to the cores of the clusters.}
\begin{tabular}{lrrrrr}
  Source & $\Delta$ (\arcsec) & $\Delta/\Rc$ & \Lx\ ($\times 10^{32}$) &
  $\cal{L}$ (\%) & $P_{\rmn{GC}}$ (\%)\\ \hline
  M92C & $17.\pm 3.$ & $1.2 \pm 0.2$ & $3.2 \pm 0.9$ & 28 & 99 \\
  M13G & $37.\pm 5.$ & $0.7 \pm 0.1$ & $6.5 \pm 1.3$ &  7 & 96 \\ \hline
\end{tabular}
\medskip 

$\Delta$ is the distance of the source from the cluster centre, $\Rc$
is the cluster core radius of 14.1 arcsec for M92 and 52.5 arcsec for
M13 (Trager et al.\ 1993) and \Lx\ is the source X-ray luminosity
(0.1--2.4 keV) in units of \ergsec.  The single-source likelihood
$\cal{L}$ is the confidence with which we prefer a single to a
double-source fit (see text) -- the quoted value for M92C refers to
the first observation only, because of aspecting problems in the
second.  $P_{\rmn{GC}}$ is the probability that the source is
associated with its globular cluster, given a source system mass of
1.4\Msol\ (see text).
\label{tbl:core}
\end{table}
of association of 99 per cent for source M92C, and 96 per cent for
source M13G.

If the mass of the low-luminosity source systems is not 1.4\Msol, this
will alter the expected distribution of the sources, and the
probability of association, accordingly.  If the systems are more
massive than 1.4\Msol, the probabilities are not much affected.  If
the systems are less massive, however, this reduces the probability of
association for both of our sources, and for M13G especially.  For
example, if the system mass is 0.7\Msol\ ($q=1.2$), our probabilities
of association are 98 per cent for M92C and 88 per cent for M13G.

Using our duplicate observation of M92, we are able to test for
long-term variability of the source M92C (Table~\ref{tbl:m92var}).  We
find no evidence for a change in the source's flux from the initial
observation to the one that occurred five months later; our statistics
are sufficient, however, only to detect variability by a factor of
three or more.

We test for multiplicity of these sources by performing a
multiple-source fit on the data sets and comparing the double-source
and single-source likelihoods by the usual $\chi^2$ prescription.
Tests on simulated data sets indicate that, for two adjacent sources
each half the strength of our M92C, we can distinguish the two-source
nature of the profile at a separation of five arcsec or more.  With
our actual data, we find that a double-source fit is not preferred for
M92C in our first M92 observation.  Examination of data from the
second observation reveals that it is probably affected by elliptical
blurring due to inaccurate aspect correction as discussed in the HRI
Report \cite{Davidet95}: the two on-axis sources M92B and M92C exhibit
an ellipticity of profile that is similar in magnitude (major and
minor axes of 7 arcsec and 4 arcsec) and orientation.  Because of
this, we do not attempt a single versus double-source comparison for
this observation.

For M13G we find that a two-source fit is mildly preferable (93 per
cent confidence level).  The magnitude and orientation of M13G's
deformation from a single-source profile is not reproduced in the
other two on-axis sources, so it does not seem to be the result of the
elliptical blurring mentioned above.  If it is in fact due to a second
source, then the two components of M13G have fluxes roughly in the
ratio of 3:1 and are separated by 6.5 arcsec.  At the distance of the
cluster, then, their X-ray luminosities would be $4.7 \times 10^{32}$
and $1.6 \times 10^{32}$ \ergsec, respectively, and they would be
separated by 0.2~pc in projection.


\section{Discussion: Excess Emission}
\label{sec:extend}

Multiple low-luminosity sources have now been seen in the cores of
several globular clusters (Hasinger et al.\ 1994; Cool et al.\ 1993;
Grindlay 1994), and the weakest detected sources have luminosities
well below our threshold.  Thus we would not be surprised if there
were more sources still undetected in the cores of M92 and M13.

\begin{figure}
 \begin{center}
   ~\psfig{figure=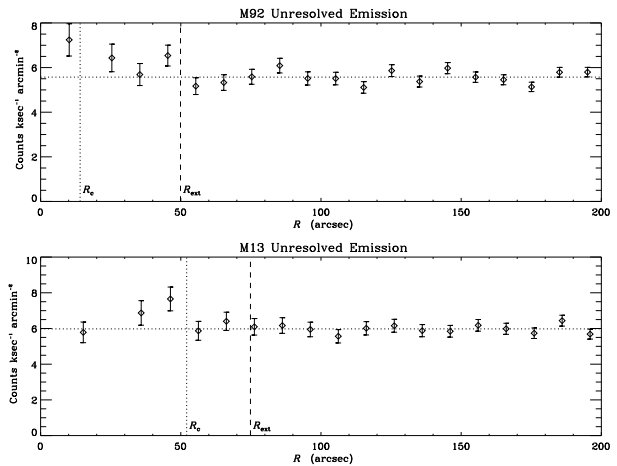,bbllx=62pt,bblly=62pt,bburx=554pt,bbury=735pt,angle=90,width=8.3cm}~
 \end{center}
 \caption{Evidence for excess emission from the cluster cores.  The
   background, determined from counts in the annulus from 2--3.3
   arcmin, is shown as a horizontal dotted line.  Points are plotted
   at the midpoint of their bins, with Poisson errors.  The dotted
   vertical line indicates the size of the cluster core \Rc, and the
   dashed vertical line indicates the rough extent of the excess
   emission, \Rext.  The emission in M13, localized near the edge of
   the core, is significantly larger than that expected from the
   ``bump'' near our source M13G (see Fig.~\protect\ref{fig:m13core}).
   }
 \label{fig:extend}
\end{figure}

We searched for the excess emission associated with such sources by
extracting photons from a circular region extending out to several
arcminutes from the centres of the clusters.  Emission from each of
our detected sources was punched out with a 12~arcsec-radius disk --
removing 90 per cent of its flux -- and replaced with the events in a
similar disk at the same radial distance from the cluster centre.  We
created plots of the binned counts in successive annuli about the
cluster centre, collecting a minimum of 100 counts in each bin, to
determine the rough extent of any excess emission
(Fig.~\ref{fig:extend}).  Then the
evidence for excess emission was simply the Poisson probability of the
corresponding fluctuation (Table~\ref{tbl:extend}).  
\begin{table}
\caption{Evidence for excess emission from the clusters.  }
\begin{tabular}{lrrrr}
  Cluster & \Rext\ (\arcsec) & (pc) & \Pext\ (\%) & \Lx\ ($\times 10^{32}$) \\
  \hline
  M92  &  50  &   1.9  &  99.8    & $4.4 \pm 1.6$  \\
  M13  &  75  &   2.7  &  98.\hspace*{0.5em} & $5.7 \pm 2.9$  \\ \hline
\end{tabular}
\medskip

\Rext\ is the rough extent of the emission, and \Pext\ is the
confidence level of its existence.  Luminosities (0.1--2.4 keV) are in
\ergsec, assuming a 1~keV bremsstrahlung spectrum (i.e., a set of
sources like the known low-luminosity sources).
\label{tbl:extend}
\end{table}

Di Stefano \& Rappaport (1994) have performed simulations of tidal
interactions in the cores of the globular clusters 47~Tuc and
$\omega$~Cen, with the aim of predicting the populations of
cataclysmic variables expected to form within them.  Performing a
simple-minded scaling of their results to our clusters, we find that
the $4.4 \times 10^{32}$ \ergsec\ of emission from M92 can correspond
to a population of roughly 30 CVs in their scheme, and that the $5.7
\times 10^{32}$ \ergsec\ of emission from M13 -- if it is real --
can correspond to a population of roughly 40 CVs.

On a more observational level, we may consider the best-guess
luminosity function for the low-luminosity sources, derived by
Johnston \& Verbunt (1996) from the full set of globular cluster
observations.  Assuming a power-law form ($dN \sim L^{-\gamma}\,
d\,\ln L$), Johnston \& Verbunt find a best $\gamma$ of about 0.5.
According to this line of reasoning, then, the emission remaining
after subtraction of the detected sources is probably due to a small
number of point sources close below our detection limit.
Figure~\ref{fig:extend} lends support to this view, since most of the
excess emission is found in a narrow radial range.  We note that the
presence of a small number of sources at these luminosities would not
necessarily contradict the predictions of a much larger number of CVs
made by DiStefano and Rappaport (see above), since the luminosity
distribution of these systems is not well known, and is compatible
with a low luminosity for most (for a recent review see Verbunt 1996).


\section{Conclusions}

We have observed a low-luminosity X-ray source in close proximity to
the core of each of the globular clusters M92 (NGC 6341) and M13 (NGC
6205).  In each case the positional coincidence indicates that the
source can probably, but not conclusively, be associated with its
cluster (probabilites of 99 per cent and 96 per cent for M92C and
M13G, respectively).  We find evidence that the source M13G has a
secondary or extended component (93 per cent).  We find no evidence
for variability in source M92C from one of our 20~ksec observations of
the cluster to the next, five months later; however, our sensitivity
is sufficient only to detect variability by a factor of three or more.

We observe excess X-ray emission from M92 (99.8 per cent confidence),
and find evidence for similar emission, unassociated with any of our
detected sources, from M13 (98 per cent confidence).  In M92 it
extends out to a distance of roughly two parsecs from the cluster
centre, and has a total luminosity (assuming a 1~keV bremsstrahlung
spectrum) of $4.4 \pm 1.6 \times 10^{32}$ \ergsec.  In M13 the excess
emission, if it exists, has a luminosity of $5.7 \pm 2.9 \times
10^{32}$ \ergsec and is localized near the edge of the core.  In both
cases, our preferred interpretation of the emission is as a small set
of sub-threshold (low-luminosity) X-ray sources within the globular
clusters.  Such a population of low-luminosity sources is predicted to
exist in the clusters by tidal capture theory (which holds that they
are CVs) and by the power-law luminosity function that best fits the
full population as currently known.

We find that all five of the X-ray sources within the M92 field of
view, but external to the cluster, are coincident with foreground 
stars from the Simbad database.

\subsection{Acknowledgments}
We thank Eric Deutsch for supplying the SASS-output positions and
count rates for the sources from the PSPC observations of the
clusters.  This material is based upon work supported under a National
Science Foundation Graduate Research Fellowship (D.~Fox).  B.~Margon
acknowledges the financial support of NASA Grant NAG5-1518.
F.~Verbunt is supported by the Netherlands Organization for Scientific
Research (NWO) under grant PGS 78-277.  The Digitized Sky Surveys were
produced at the Space Telescope Science Institute under U.S.
Government grant NAG W-2166.  This research made use of the Simbad
database, operated at CDS, Strasbourg, France.


\label{lastpage}

\end{document}